\documentstyle[12pt]{article}

\newcommand{\be}{\begin{equation}}
\newcommand{\ee}{\end{equation}}
\newcommand{\bea}{\begin{eqnarray}}
\newcommand{\eea}{\end{eqnarray}}

\begin{document}
\begin{titlepage}
\rightline{To Appear: {\em Class. Quantum Grav.}}

\vspace{1in}

\begin{center}
\Large
{\bf Symmetric vacuum scalar--tensor cosmology}

\vspace{1in}

\normalsize

\large{James E. Lidsey$^1$}

\normalsize
\vspace{.7in}

{\em Astronomy Unit, School of Mathematical 
Sciences,  \\ 
Queen Mary \& Westfield, Mile End Road, LONDON, E1 4NS, U.K.}

\end{center}

\vspace{1in}

\baselineskip=24pt
\begin{abstract}
\noindent
The existence of point symmetries in the cosmological field equations 
of generalized vacuum scalar--tensor theories is 
considered within the context of the spatially homogeneous 
cosmologies. It is found that such symmetries only occur in the   
 Brans--Dicke theory when the dilaton field 
self--interacts. Moreover, the interaction potential of the dilaton
must take the form of a cosmological constant. 
For the spatially flat, isotropic model, it is shown how 
this point symmetry may be employed to generate a discrete 
scale factor duality in the Brans--Dicke action. 

\end{abstract}

\vspace{.7in}

 
\vspace{.2in}
$^1$Electronic address: jel@maths.qmw.ac.uk

\end{titlepage}


In this paper we search for point symmetries in the 
cosmological field equations of  
generalized vacuum scalar--tensor theories of gravity. 
Interest in these theories has been widespread in recent years.
They are defined by the action 
\be 
\label{action}
S=\int d^4 x \sqrt{-g} e^{-\Phi}
 \left[  R - \omega (\Phi ) \left( 
\nabla \Phi \right)^2 -2\Lambda (\Phi) \right] ,
\ee
where $R$ is the Ricci curvature of the space--time and 
$g$ is the determinant of the 
metric $g_{\mu\nu}$ \cite{ST}. The dilaton field $\Phi$ 
plays the role of a time--varying gravitational constant
and may self--interact through a potential  $\Lambda (\Phi )$. 
The function $\omega (\Phi)$ is dimensionless and 
determines the precise form of the coupling between the dilaton and graviton.
Each scalar--tensor theory is defined by the functional forms of 
$\omega (\Phi)$ and $\Lambda (\Phi)$. A cosmological constant in the 
gravitational sector of the theory corresponds to 
the special case where $\Lambda (\Phi)$ is a space--time constant. 

Action (\ref{action}) provides a 
natural background within which deviations from general relativity 
may be quantitatively studied. The simplest example  is the 
Brans--Dicke theory, where $\omega (\Phi)$ is a space--time constant 
\cite{BD}.
It is known that inflationary solutions exist  in a wide class of 
scalar--tensor cosmologies and these theories are therefore
relevant to the study of the very early Universe \cite{INF}. 
Indeed, higher--order \cite{W} and higher--dimensional \cite{HD}
theories of gravity may be expressed in a scalar--tensor 
form after suitable field redefinitions and the Brans--Dicke theory 
with $\omega =-1$ corresponds to a truncated version 
of the string effective action \cite{STRING}. 

Point symmetries associated with action (\ref{action}) 
have been discussed previously within the 
context of the spatially isotropic Friedmann Universes \cite{PS,PS2}. 
It was found that $\omega (\Phi)$ and $\Lambda (\Phi)$ 
must be related in a  certain way if the field equations are to be symmetric. 
In this paper we consider whether theory (\ref{action}) 
admits point symmetries for the more general class of spatially homogeneous
Bianchi Universes. We assume that $\Lambda (\Phi ) \ne 0$ and 
that $\omega (\Phi ) >-3/2$ for all physical values of $\Phi$. We find that 
such symmetries only exist in these anisotropic cosmologies if strong 
restrictions are imposed on the form of Eq. (\ref{action}). In particular, 
we show that for the Bianchi type I 
model, $\omega (\Phi)$ and $\Lambda (\Phi)$ must both be {\em constant}. 
We argue that this conclusion should apply 
for the other Bianchi types where a Lagrangian 
formulation of the field equations is possible. 

The line element for the class of spatially homogeneous space-times is given by
\be
\label{metric}
ds^2=-dt^2 +h_{ab} \omega^a \omega^b, \qquad a,b=1,2,3 ,
\ee
where $h_{ab}(t)$ is a function of cosmic time $t$ and 
represents the metric on the surfaces of homogeneity and
$\omega^a$ are one--forms. These models have 
a topology $R\times G_3$, where $G_3$ represents a Lie group of isometries 
that acts transitively on the space--like three--dimensional orbits 
\cite{RS}. The Lie algebra of $G_3$ admits 
the structure  constants ${C^a}_{bc}=m^{ad}\epsilon_{dbc}+
{\delta^a}_{[b} a_{c]}$, where $m^{ab}$ is a symmetric matrix, $a_c 
\equiv {C^a}_{ac}$ and $\epsilon_{abc} = \epsilon_{[abc]}$. 
The Jacobi identity 
${C^a}_{b[c} {C^b}_{de]} =0$ is only satisfied if $m^{ab}a_b =0$, so 
$m^{ab}$ must be transverse to $a_b$ \cite{10}. The model belongs to the 
Bianchi class A if $a_b=0$ and to the class B if $a_b \ne 0$.
A basis may be found such that $a_b =(a,0,0)$ and 
$m^{ab} ={\rm diag} \left[ m_{11},m_{22},m_{33} \right]$, 
where $m_{ii}$ take the values  $\pm 1$ or $0$.
In the Bianchi class A, 
the Lie algebra is uniquely determined up to isomorphisms 
by the rank and signature of $m^{ab}$. The six possibilities are 
$(0,0,0)$, $(1,0,0)$, $(1,-1,0)$, $(1,1,0)$, $(1,1,-1)$ and $(1,1,1)$ and  
these correspond, respectively, to the Bianchi types I, II, ${\rm VI}_0$, 
${\rm VII}_0$, VIII and IX. Finally, 
the three-metric may be parametrized by $h_{ab}(t)=e^{2\alpha (t)} 
\left( e^{2\beta (t)} \right)_{ab}$, where $e^{3\alpha}$ represents the 
effective spatial volume of the Universe and 
\be
\beta_{ab} \equiv 
{\rm diag} \left[ \beta_+ +\sqrt{3}\beta_-,\beta_+ -\sqrt{3}\beta_- , -2
\beta_+ \right]
\ee
is a traceless matrix that determines the anisotropy in the models. 

The configuration space $Q$ for the Bianchi  
models derived from action (\ref{action}) is 
therefore four--dimensional and is 
spanned by  $\{ q_n \equiv \alpha , \Phi ,\beta_{\pm} \}$. 
The Lagrangian density $L (q_n ,\dot{q}_n )$ is  
defined by $S=\int dt L (q_n , \dot{q}_n )$, 
where a dot denotes differentiation 
with respect to cosmic time. It may be derived by substituting 
the {\em ansatz} (\ref{metric}) into action (\ref{action}) and 
integrating over the spatial variables. This procedure is unambiguous
for the class A cosmologies and the action for these models simplifies to 
\be
\label{actionbianchi}
S = \int dt e^{3\alpha -\Phi} \left[ 
6\dot{\alpha} \dot{\Phi} -6\dot{\alpha}^2 +6\dot{\beta}^2_+ +
6\dot{\beta}^2_- +\omega (\Phi) 
\dot{\Phi}^2  -2\Lambda (\Phi ) +e^{-2\alpha} U(\beta_{\pm})  
\right]    ,
\ee
where 
\be
\label{potentialA}
U(\beta_{\pm}) = -e^{-4 \alpha} \left( m_{ab} m^{ab} -\frac{1}{2} m^2 
\right) 
\ee
is the curvature potential, 
$m \equiv {m^a}_a$ and indices are raised and lowered with $h^{ab}$ 
and $h_{ab}$, respectively \cite{WALD}. In the case of the 
type B models, a divergence may arise because the three--curvature 
contains a term proportional to $a_b a^b$  \cite{Mac}. In view of this, 
we do not consider these models further.

The field equations derived from action (\ref{actionbianchi}) take the 
familiar  form
\be
\label{EL}
\frac{d}{dt} \frac{\partial L}{\partial \dot{q}_n} = 
\frac{\partial L}{\partial q_n}  .
\ee
Now, a point symmetry of a set of differential 
equations such as those given by Eq. (\ref{EL}) 
may be viewed as a one--parameter 
group of transformations acting in the space $TQ$ 
that is tangent to $Q$ and spanned 
by $\{q_n ,\dot{q}_n \}$. One identifies such a symmetry 
by introducing a set of arbitrary, real, 
differentiable  functions $\{ X_n(q) 
=X_{\alpha}(q), X_{\Phi}(q), X_{\pm}(q) \}$ and  
contracting these with the field equations \cite{PS1}. Thus, 
\be
\label{equation}
\frac{d}{dt} \left( X_n \frac{\partial L}{\partial \dot{q}_n} \right) 
= \left( X_n \frac{\partial}{\partial q_n} +\frac{dX_n}{dt} 
\frac{\partial}{\partial \dot{q}_n} \right) L ,
\ee
where summation over $n$ is implied. The right hand side 
of this equation is the Lie derivative ${\cal{L}}_{\bf X} L$
of the Lagrangian density   with respect to the vector field 
\be
\label{vector}
{\rm {\bf X}} \equiv  
X_n \frac{\partial}{\partial q_n} +\frac{dX_n}{dt} 
\frac{\partial}{\partial \dot{q}_n} .
\ee
This vector field belongs to the tangent space and is the infinitesimal 
generator of the point transformation. The 
Lie derivative (\ref{equation}) determines how the Lagrangian
varies along the flow generated by ${\bf X}$ in $TQ$. 
When this derivative 
vanishes, the Lagrangian density is {\em constant} along the integral curves 
of ${\rm {\bf X}}$. It then follows immediately from 
Eq. (\ref{equation}) that the quantity 
$i_{\bf {X}} \theta_{L} \equiv  X_n {\partial L}/{\partial \dot{q}_n}$ 
is conserved,
where $i_{\bf {X}}$ denotes the contraction of the vector field ${\bf X}$ 
with $\theta_L \equiv (\partial L /\partial \dot{q}_n)dq_n$.

Thus, one may uncover a Noether--type 
symmetry in the theory by determining the components $\{ X_n (q) \}$ that
satisfy ${\cal{L}}_{\bf X} L=0$ \cite{PS1}. 
In general, this equation reduces to 
an expression that is quadratic in $\dot{q}_n$ for all values of $n$. 
However, the coefficients of these terms are determined 
by functions of $q_n$. 
Thus, each of the coefficients must vanish identically if 
the Lie derivative is to vanish. 
This leads to a number of 
separate constraints that take the form of first--order,  partial 
differential equations in $\{ X_n (q) \}$. A further constraint may 
arise from terms in the Lagrangian that are independent of $\dot{q}_n$. 
A Noether symmetry in the theory 
is then identified once a solution to these equations is found.

It can be shown after some algebra that 
the Lie derivative of the Lagrangian (\ref{actionbianchi}) 
with respect to ${\bf X}$ vanishes
if and only if $\{ X_n (q) \}$ satisfy the set of partial differential 
equations:
\bea
\label{1}
6\Lambda X_{\alpha} -2\Lambda X_{\Phi} +2\Lambda 'X_{\Phi} =
e^{-2\alpha} \left[ X_{\alpha} U -X_{\Phi} U +X_+ \frac{\partial U}{\partial 
\beta_+} +X_- \frac{\partial U}{\partial \beta_-} \right] \\
\label{3}
9X_{\alpha} -3X_{\Phi} +3 \frac{\partial X_{\alpha}}{\partial \alpha}
-6\frac{\partial X_{\alpha}}{\partial \Phi} +3\frac{\partial 
X_{\Phi}}{\partial \Phi} +\omega \frac{\partial X_{\Phi}}{\partial \alpha} 
=0 \\
\label{4}
3\omega X_{\alpha} -\omega X_{\Phi} +\omega' X_{\Phi} +6 \frac{\partial 
X_{\alpha}}{\partial \Phi}
+2\omega \frac{\partial X_{\Phi}}{\partial \Phi} =0 \\
\label{5}
3X_{\alpha}-X_{\Phi} +2\frac{\partial X_{\alpha}}{\partial \alpha} 
-\frac{\partial X_{\Phi}}{\partial \alpha} =0 \\
\label{6}
3X_{\alpha} -X_{\Phi} +2\frac{\partial X_{\pm}}{\partial \beta_{\pm}} =0  \\
\label{7}
\frac{\partial X_+}{\partial \beta_-} +\frac{\partial X_-}{\partial 
\beta_+} =0 \\
\label{8}
3\frac{\partial X_{\alpha}}{\partial \beta_{\pm}} +\omega \frac{\partial 
X_{\Phi}}{\partial \beta_{\pm}} +
6 \frac{\partial X_{\pm}}{\partial \Phi} =0 \\
\label{9}
-2 \frac{\partial X_{\alpha}}{\partial \beta_{\pm}} +\frac{\partial 
X_{\Phi}}{\partial \beta_{\pm}} +2\frac{\partial X_{\pm}}{\partial 
\alpha} =0  ,
\eea
where a prime denotes differentiation with respect to $\Phi$. 

We will search for non--trivial solutions to these equations
where  $\{ X_{\alpha} ,X_{\Phi} , \Lambda (\Phi ) \ne 0 \}$. Moreover, 
we shall consider the case where both 
sides of Eq. (\ref{1}) are identically zero:
\bea
\label{2a}
X_{\alpha} = \left( \frac{\Lambda -\Lambda '}{3\Lambda} 
\right) X_{\Phi} \\
\label{2b}
X_{\alpha} U -X_{\Phi} U +X_+ \frac{\partial U}{\partial 
\beta_+} +X_- \frac{\partial U}{\partial \beta_-} =0  .
\eea
This separation is valid in general for the type I model, since the curvature 
potential $U(\beta_{\pm})$ is identically zero in this case. However, 
it should also be consistent for the 
other Bianchi types. Eq. (\ref{6}) implies that 
$\partial X_+/\partial \beta_+ =\partial X_- /\partial \beta_-$. 
If we differentiate this constraint 
with respect to $\beta_{\pm}$ and compare it with the first 
derivative of Eq. (\ref{7}), we find that $X_{\pm}$ satisfy the 
one--dimensional Laplace equation:
\be
\label{laplace}
\frac{\partial^2 X_{\pm}}{\partial \beta_{\pm}^2} + 
\frac{\partial^2 X_{\pm}}{\partial \beta_{\mp}^2} =0  .
\ee
Now, the components of ${\bf X}$ must be real if they 
are to correspond to physical solutions. 
However, an exponential solution to Eq. (\ref{laplace})
will have the generic form $X_j = \exp \left[ ik \beta_{\pm} \pm 
k\beta_{\mp} \right]$, for some arbitrary, real constant $k$. 
This suggests that 
$X_{\pm}$ can not contain real exponential terms in $\beta_{\pm}$. 
Furthermore, Eqs. (\ref{8}) and (\ref{9}) then imply that 
the same will be 
true for $X_{\alpha}$ and $X_{\Phi}$. This is important because 
the curvature 
potential (\ref{potentialA}) consists entirely of 
exponential terms. We might expect, therefore, that 
the components of ${\bf X}$ will be unable to cancel out these 
terms in the full expression given by 
Eq. (\ref{1}). If so, Eq. (\ref{1}) could only 
be satisfied if both sides were identically zero. 

When Eq. (\ref{2a}) is valid, Eq. (\ref{5}) simplifies to
\be
\label{Xalpha}
\frac{\partial \ln X_{\Phi}}{\partial \alpha} = c(\Phi) = - 
\left( \frac{3\Lambda '}{\Lambda + 2\Lambda '} \right) .
\ee
On the other hand, 
we may combine Eqs. (\ref{3}) and (\ref{4}) and 
eliminate the $\partial 
X_{\alpha} / \partial \Phi$ terms. Substituting Eq. (\ref{Xalpha}) 
into the result then implies that 
\be
\label{XPhi}
\frac{\partial \ln X_{\Phi}}{\partial \Phi} = f(\Phi) =  
\frac{1}{3+2\omega} 
\left[ (3+\omega )\frac{\Lambda '}{\Lambda} + \frac{3\Lambda '}{\Lambda 
+ 2\Lambda'} \left( 1+\omega -\frac{\Lambda '}{\Lambda} \right) -
\omega' \right]  .
\ee
It follows from Eqs. (\ref{Xalpha}) and (\ref{XPhi}) that 
$X_{\Phi}$ must be separable  in
$\alpha$ and $\Phi$. Inserting a separable ansatz into Eq. (\ref{Xalpha}) 
then implies that $c(\Phi) \equiv c$ must be {\em independent} 
of $\Phi$. 

We may also equate Eqs. (\ref{5}) and (\ref{6}) and differentiate 
with respect to $\alpha$. The term containing second derivatives in $X_{\pm}$ 
may then be eliminated by substituting the differential of Eq. (\ref{9}) with 
respect to $\beta_{\pm}$. Moreover, substitution of Eq. (\ref{2a}) then 
removes any direct dependence on $X_{\alpha}$. 
This procedure leads to the very useful constraints
\be
\label{second}
\frac{\partial^2 X_{\Phi}}{\partial \alpha^2} = \frac{\partial^2 
X_{\Phi}}{\partial \beta^2_+} = \frac{\partial^2 X_{\Phi}}{\partial 
\beta_-^2} = c^2 X_{\Phi}
\ee
on the second derivatives of $X_{\Phi}$.  These derivatives may be 
related to those of $X_{\pm}$ by rewriting Eq. (\ref{6}):
\be
\label{6a}
\frac{\partial X_{\pm}}{\partial \beta_{\pm}} = -\frac{c}{4c+6} X_{\Phi}  .
\ee
If we differentiate this equation twice with respect to $\beta_{\pm}$, we 
find that 
\be
\label{twiceplus}
\frac{\partial^3 X_{\pm}}{\partial \beta_{\pm}^3} = -\frac{c^3}{4c+6} 
X_{\Phi}
\ee
after substitution of Eq. (\ref{second}). 
Differentiating  Eq. (\ref{6a}) twice with 
respect to $\beta_{\mp}$ then implies that
\be
\label{twiceminus}
\frac{\partial^3 X_{\pm}}{\partial \beta^2_{\mp} \partial \beta_{\pm}}
=-\frac{c^3}{4c+6} X_{\Phi}  .
\ee

However, differentiation of Eq. (\ref{laplace}) with respect to 
$\beta_{\pm}$ implies that
\be
\label{pm}
\frac{\partial^3 X_{\pm}}{\partial \beta^3_{\pm}} = - 
\frac{\partial^3 X_{\pm}}{\partial \beta^2_{\mp} \partial \beta_{\pm}}  ,
\ee
so Eqs. (\ref{twiceplus}), (\ref{twiceminus}) and (\ref{pm})  
are only consistent if $c=0$. Thus, 
$\Lambda (\Phi)$ must be a space--time constant.
Eq. (\ref{2a}) then implies that 
$3X_{\alpha}=X_{\Phi}$ if $\Lambda \ne 0$. When this condition is satisfied, 
Eq. (\ref{5}) implies that $X_{\alpha}$ and $X_{\Phi}$ must be independent 
of $\alpha$, as expected. It then follows from Eq. (\ref{3}) that 
these functions must also be  independent of $\Phi$. Moreover, 
Eq. (\ref{4}) can only be satisfied in this case if $\omega' =0$. Thus,  
$\omega (\Phi)$ must also be a space--time constant and this corresponds to 
the Brans--Dicke theory. 

The solution to Eqs. (\ref{6})--(\ref{9})  for constant $\omega$ 
is found to be
\bea
\label{gensol}
X_{\Phi} =h_0 +h_+ \beta_+ +h_-\beta_- ,\qquad X_{\Phi} =3X_{\alpha} 
\nonumber \\
X_{\pm} = x_{\pm} +b_{\pm} \beta_{\mp} -\frac{h_{\pm}}{6} (1+\omega ) 
\Phi -\frac{h_{\pm}}{6} \alpha    ,
\eea
where $\{ h_0 , b_+ , h_{\pm} ,x_{\pm} \}$ are arbitrary 
constants and $b_-=-b_+$. However, Eq. (\ref{2b}) must also be solved. 
This condition is trivial for the type I 
model,  but it places further restrictions on the 
components of ${\bf X}$ in the case of the other Bianchi types. 
Since there are no exponential terms in 
Eq. (\ref{gensol}), Eq. (\ref{2b}) 
must reduce to six separate constraints:  
\bea
\label{m}
m_{11}^2 \left[ X_{\alpha}-2X_+ -2\sqrt{3} X_- \right] =0 \nonumber \\
m_{22}^2 \left[ X_{\alpha}-2X_+ +2\sqrt{3} X_- \right] =0 \nonumber \\
m_{33}^2 \left[ X_{\alpha} +4 X_+ \right] =0 \nonumber \\
m_{11} m_{22} \left[ X_{\alpha} -2X_+ \right] =0 \nonumber \\
m_{11}m_{33} \left[ X_{\alpha}  +X_+ -\sqrt{3} X_- \right] =0 \nonumber \\
m_{22}m_{33} \left[ X_{\alpha} +X_+ +\sqrt{3} X_- \right] =0   .
\eea
In the case of the type II model, these equations are satisfied when 
$h_0=6(x_+ +\sqrt{3} x_-)$ and $h_+=-\sqrt{3}h_- = -6\sqrt{3} b_+$. 
For types ${\rm VI}_0$ and ${\rm VII}_0$, the stronger restrictions 
$X_-=0$ and $X_{\alpha}=2X_+={\rm constant}$ must apply if a symmetry 
is to exist. However, the only solution to Eq. (\ref{m}) 
for types VIII and IX is $\{ X_n \} =0$, so the field equations 
for these two models do not admit non--trivial point symmetries. 

Eq. (\ref{gensol}) is 
the general solution to Eqs. (\ref{3})--(\ref{9}) when 
$\Lambda (\Phi) \ne 0$ and Eqs. (\ref{2a}) and (\ref{2b}) are valid. 
When these conditions are satisfied, therefore, we may conclude that 
the only vacuum scalar--tensor gravity theory that contains a point symmetry 
in anisotropic cosmologies is the Brans--Dicke theory with a 
cosmological constant in the gravitational sector of the theory. 

The Brans--Dicke theory also exhibits 
this symmetry when $\Lambda =0$. This may be verified by 
substituting the ansatz $X_{\Phi} (\beta_{\pm}) =3X_{\alpha} (\beta_{\pm})$ 
into Eqs. (\ref{1})--(\ref{9}). We should emphasize, however, that 
other theories may also be symmetric when $\Lambda$ vanishes since the 
left hand side of Eq. 
(\ref{1}) is trivial in this case. Consequently, Eq. (\ref{2a}) does not 
apply, so 
Eqs. (\ref{Xalpha}) and (\ref{XPhi}) are not the unique solutions 
to Eqs. (\ref{3})--(\ref{5}). This implies that $X_{\Phi}$ 
could take a more general form to that given in Eq. (\ref{gensol}). 
It would be of interest to investigate whether other 
theories are indeed symmetric when the dilaton potential vanishes. 

Recently, a further symmetry in the Brans--Dicke cosmology was identified 
within the context of the spatially 
flat, isotropic Friedmann Universe \cite{LIDS}. 
It can be shown by direct substitution 
that action (\ref{actionbianchi}) is invariant under 
a scale factor duality transformation 
\bea
\label{sfd}
\alpha =\frac{2+3\omega}{4+3\omega} z - \frac{2(1+\omega)}{4+3\omega} w
\nonumber \\
\Phi = -\frac{6}{4+3\omega} z-\frac{2+3\omega}{4+3\omega} w
\eea
when $\omega \ne -4/3$ and $U=\beta_{\pm} =\Lambda '=0$. 
This symmetry is a generalization of the 
scale factor duality exhibited by the string effective action \cite{duality}.
It is a discrete symmetry but 
it may be related to the continuous Noether symmetry discussed in this 
work. 

In the isotropic case the configuration space 
is two--dimensional and the Lie derivative of the Lagrangian 
vanishes if $X_{\Phi} =3X_{\alpha} ={\rm constant}$ \cite{PS2}. 
This Noether symmetry may be employed to generate a new set of variables
$q_n =q_n (Q_k)$ $(n,k =1,2)$. In this case 
the vector field (\ref{vector}) transforms to 
\be
{\bf X} =\left( i_{\rm {\bf X}} dQ_k \right) 
\frac{\partial}{\partial Q_k} + \left[ \frac{d}{dt} 
\left( i_{\rm {\bf X}} dQ_k \right) \right] \frac{\partial}{\partial 
\dot{Q}_k}   ,
\ee
where the contraction is over ${\rm {\bf X}}$ and $dQ_k =(\partial 
Q_k/\partial q_n) dq_n$ \cite{foliation}. 
We may define   $\{ Q_k \} $ such that they satisfy 
the first--order partial differential equations 
\be
\label{Qconstraint}
i_{\rm {\bf X}} dw =\epsilon_1(w,z) , 
\qquad i_{\rm {\bf X}} dz  =\epsilon_2(w,z)    ,
\ee
where $w \equiv Q_1$, $z\equiv Q_2$ and 
$\epsilon_l (w,z)$ are particular functions. If we specify 
these functions as 
$\epsilon_1 =-3 X_{\alpha}$ and $\epsilon_2 
=-X_{\alpha}$, respectively, the  solution to Eq. (\ref{Qconstraint}) 
is given by Eq. (\ref{sfd}). Thus, the scale factor 
duality of the Brans--Dicke theory may be generated by 
the point symmetry associated with the vector field ${\bf X}$. 

We will conclude with some general remarks. The symmetry discussed in this 
work has a number of applications. Firstly, it 
leads to a conserved quantity of the form 
$i_{\bf X} \theta_L = X_n  \partial L/ \partial 
\dot{q}_n $. Since the Lagrangian density is quadratic in 
$\dot{q}_n$, the conservation of $i_{\bf X} \theta_L$ results 
in an equation that relates the first derivatives of the configuration 
space variables $q_n$. This represents a first integral of the 
field equations (\ref{EL}). In principle, it should be easier to 
solve this constraint, together with the Hamiltonian constraint, 
rather than the full system given by Eq. (\ref{EL}). 

Thus, the existence of a conservation law implies that 
the field equations may be simplified considerably and this may
lead to new solutions. In particular, it would 
be interesting to derive new inflationary solutions by this 
approach. The search for exact inflationary solutions 
in anisotropic cosmologies is well motivated. These solutions would 
provide insight into how the anisotropy is effectively washed away 
by the accelerated expansion, thereby leading to 
the highly isotropic Universe that is observed today. 
The question of how the anisotropy may influence the onset of inflation
may also be addressed through exact solutions. Such solutions will exist
since the symmetry is compatible with a cosmological 
constant in the gravitational sector of the theory. This term 
could also arise, for example, 
from the potential energy of a second scalar field that is coupled 
to the dilaton field in an appropriate fashion. 

We have shown that a Noether symmetry 
arises in the  Bianchi types I, II, ${\rm VI}_0$ and ${\rm VII}_0$ and have
argued that it is unique  to the Brans--Dicke theory 
in these cases. However, our conclusions
also apply to other Bianchi models. Although a symmetry of the 
form discussed here does not exist for the Bianchi types VIII and IX, 
we may consider the high anisotropic limit of all 
Bianchi A models where $\beta_{\pm} \gg 1$. In this case, $h_{22}/h_{11} 
\ll 1$ and $h_{33}/h_{11} \ll 1$, so 
the dominate term in the curvature potential
(\ref{potentialA}) is $m_{11}^2h_{11}^2$. 
In effect, this is equivalent to specifying $m_{11}=1$ and $m_{22}
=m_{33}=0$ in Eq. (\ref{potentialA}) and this corresponds to 
the type II model. Hence,  
the field equations of the Bianchi types VIII and IX will exhibit an 
approximate point symmetry if the anisotropy  is sufficiently 
large. This is interesting because the initial state of the Universe
may well have been very anisotropic due to 
quantum effects and the early Universe
is precisely the regime where scalar--tensor gravity is thought 
to  have been relevant. 

We have not considered the Bianchi 
class B models directly 
in this work because the Lagrangian description of the field 
equations is not always consistent \cite{Mac}. However, our conclusions will 
apply for those models in this class 
that can be expressed in a Lagrangian form, 
because the corresponding curvature potential will contain 
exponential terms in $\beta_{\pm}$ \cite{WALD}. 
Consequently, 
the separation of Eq. (\ref{1}) into Eqs. (\ref{2a}) and (\ref{2b}) 
will apply in these cases also. Thus, the point symmetry associated with 
the Brans--Dicke theory arises in a number of different 
homogeneous cosmologies. 

The question of which scalar-tensor theory may have applied 
in the early Universe is currently unresolved. 
There are two approaches that one might take in 
addressing this question. Firstly, one may 
identify the subset of theories that are attracted to the general 
relativistic limit at late times \cite{late}. Alternatively, one may 
attempt to uncover a deeper principle that strongly favours one 
particular theory. Symmetries often provide strong motivation 
for selecting a given theory from the space of possible theories. We 
have found that the requirement that a point symmetry 
exists in the homogeneous, cosmological
 field equations of generalized scalar--tensor 
gravity is surprisingly restrictive. Indeed, the Brans--Dicke theory 
is the only theory to exhibit such a symmetry when 
the dilaton field self--interacts. This may be significant because 
the Brans--Dicke theory is 
consistent with all cosmological observations if $\omega >500$. If 
the point symmetry only arose in theories that could not 
reproduce Einstein gravity at the present epoch, it would 
be uninteresting. However, we have found that the symmetry is associated  
with a realistic theory of gravity.

Finally, we considered  
the Noether symmetry of the Brans--Dicke theory within the context of the 
spatially flat, isotropic Universe. We showed how it is 
directly related to a scale factor duality invariance of the theory. 
The Noether 
symmetry provides new insight into how the duality arises. It would 
be of interest to investigate whether the Noether symmetry 
associated with the anisotropic cosmologies may be employed in a similar 
fashion to uncover more general discrete symmetries in the Brans--Dicke 
theory. If such discrete symmetries exist, they could be employed to
map a particular solution of the field equations onto 
a new, generally inequivalent, solution. In the isotropic model, 
solutions may be generated 
in this fashion with and without a cosmological term and, indeed, 
inflationary solutions may be found that are driven entirely 
by the kinetic energy of the dilaton field \cite{levin}. A similar 
approach could be followed in the anisotropic Universes. 

To summarize, therefore, we have investigated the existence 
of point symmetries in the homogeneous, cosmological 
field equations of generalized 
vacuum scalar--tensor gravity under the assumption that the dilaton field
self--interacts. In the case of 
the spatially flat, anisotropic cosmology, we found that 
the Brans--Dicke theory containing a 
cosmological constant is the only scalar--tensor theory whose field 
equations exhibit a point symmetry. We have argued 
that this result also applies for types II, ${\rm VI}_0$ and ${\rm VII}_0$. 
We may conclude, therefore, that 
the Brans--Dicke theory exhibits
a higher level of symmetry than other scalar--tensor theories.

\vspace{.7in}

The author is supported by the Particle Physics and Astronomy 
Research Council (PPARC), UK.


\centerline{{\bf References}}
\begin{enumerate}

\bibitem{ST} Bergmann P G 1968 {\em Int. J. Theor. Phys.} {\bf 1} 25
 
Wagoner R V 1970 {\em Phys. Rev.} D {\bf 1} 3209

Nordtvedt K 1970 {\em Astrophys. J.} {\bf 161} 1059

\bibitem{BD} Brans C and Dicke R H 1961 {\em Phys. Rev.} {\bf 124} 925

\bibitem{INF} Accetta F S, Zoller D J and Turner M S 1985 {\em Phys. 
Rev.} D {\bf 32} 3046

Steinhardt P J and Accetta F S 1990 {\em Phys. Rev. 
Lett.} {\bf 64} 2470

Garc\'ia--Bellido J and Quir\'os M 1990 {\em Phys. 
Lett.} {\bf 243B} 45

Levin J J and Freese K 1993 {\em Phys. Rev.} D
{\bf47} 4282

Levin J J and Freese K 1994 {\em Nucl. Phys.} 
{\bf 421B} 635

Barrow J D and Mimoso J P 1994 {\em Phys. Rev.} D {\bf 50} 3746

\bibitem{W} Wands D 1994 {\em Class. Quantum Grav.} {\bf 11} 269

\bibitem{HD} Holman R, Kolb E W, Vadas S and Wang Y 1991 {\em Phys. 
Rev.} D {\bf 43} 995

\bibitem{STRING} Fradkin E S and Tseytlin A A 1985 {\em Nucl. Phys.} 
{\bf 261B} 1

Callan C G, Friedan D, Martinec E and Perry M J 1985 
{\em Nucl. Phys.}  {\bf 262B} 593

Green M B, Schwarz J H  and Witten E 1988 
{\em Superstring Theory} (Cambridge: Cambridge University Press)

Casas J A, Garc\'ia--Bellido J  and Quir\'os M 1991 {\em Nucl. Phys.} {\bf 
361B} 713

\bibitem{PS} Demia\'nski M, de Ritis R, Marmo G, Platania G, Rubano C, 
Scudellaro P and Stornaiolo C 1991 {\em Phys. Rev.} D {\bf 44} 3136

Capozziello S and de Ritis R 1994 {\em Class. Quantum Grav.} {\bf 11} 
107

Capozziello S, Demia\'nski M, de Ritis R and Rubano C 1995 {\em 
Phys. Rev.} D {\bf 52} 3288

\bibitem{PS2} Capozziello S and de Ritis R 1993 {\em Phys. Lett.} 
{\bf 177A} 1



\bibitem{RS} Ryan M P and Shepley L C 1975 {\em Homogeneous Relativistic 
Cosmologies} (Princeton: Princeton University Press)

\bibitem{10} Ellis G F R and MacCallum M A H 1969 {\em 
Commun. Math. Phys.} {\bf 12} 108

\bibitem{WALD} Wald R M 1983 Phys. Rev. D {\bf 28} 2118

\bibitem{Mac} MacCallum M A H 1979 in {\em General Relativity; an 
Einstein Centenary  Survey} ed. Hawking S W and Israel W 
(Cambridge: Cambridge University 
Press)

\bibitem{PS1} Marmo G, Saletan E J, Simoni A and 
Vitale B 1985 {\em Dynamical Systems} (New York: Wiley)

\bibitem{LIDS} Lidsey J E 1995 {\em Phys. Rev.} D {\bf 52} R5407

\bibitem{duality} Veneziano G 1991 {\em Phys. Lett.} {\bf 265B} 287

Gasperini M and Veneziano G 1992 {\em Phys. Lett.} {\bf 277B} 265

Tseytlin A A and Vafa C 1992 {\em Nucl. Phys.} {\bf 372B} 443

Giveon A, Porrati M and Rabinovici E 1994 {\em Phys. Rep.} {\bf 244} 177

\bibitem{foliation} Capozziello S, de Ritis R and Rubano C 1993 
{\em Phys. Lett.} {\bf 177A} 8

\bibitem{late}  Damour T and Nordtvedt  K 1993 {\em Phys. Rev.} D 
{\bf 48} 3436

Damour T and Nordtvedt  K 1993 {\em Phys. Rev. Lett.} {\bf 70} 2217


\bibitem{levin} Levin J J 1995 {\em Phys. Rev.} D {\bf 51} 462

Gasperini M and Veneziano G 1993 {\em Astropart. Phys.} {\bf 1} 317

\end{enumerate}
\end{document}